\documentstyle[prl,twocolumn,aps,psfig]{revtex}
\begin{document}                % INITIALIZE - DONT CHANGE
\def\a{$\alpha$}
\def\be{\begin{equation}}
\def\ee{\end{equation}}
\def\ba{\begin{eqnarray}}
\def\ea{\end{eqnarray}}
\draft
\preprint{DOE/ER/40561-50}
\title{Collective and chaotic motion in self-bound many-body systems}
\author{Thomas Papenbrock}
\address{Institute for Nuclear Theory, Department of Physics, 
University of Washington, Seattle, WA 98195, USA}
\maketitle
\begin{abstract}
We investigate the interplay of collective and chaotic motion in a classical
self-bound $N$-body system with two-body interactions. This system displays
a hierarchy of three well separated time scales that govern the onset of 
chaos, damping of collective motion and equilibration. Comparison with a 
mean-field problem shows that damping is mainly due to dephasing. The
Lyapunov exponent, damping and equilibration rates depend mildly on the
system size $N$.
\end{abstract}
Self-bound many-body systems like the atomic nucleus display regular,
collective phenomena as well as chaotic behavior \cite{BoWei,Zele}. The giant
dipole resonance, for example, constitutes a {\it regular} oscillation of the
protons against the neutrons albeit it is excited at energies where the
spectrum displays fluctuations which are typical for {\it chaotic} systems
\cite{GMW}. This interesting interplay of collective and chaotic motion and the
effects of chaotic dynamics on the damping and dissipation of nuclear
excitations is a matter of intense research
\cite{Koonin,Blocki78,Speth,Bertsch,Wilkinson,Abul,Blocki93,Aurel,Hofmann,Baldo,Bauer}
but still not fully understood.

Several authors \cite{Koonin,Blocki78,Abul,Blocki93,Baldo} have addressed
equilibration and damping of collective motion by coupling a slowly moving
collective degree of freedom (typically the wall of a container) to fast moving
independent particle degrees of freedom. Within these models, collective
degrees of freedom are neither constructed from single-particle coordinates nor
the result of a self-consistent mean field. In the case of driven walls this
may lead to energy conservation problems. Bauer {\it et al.} \cite{Bauer}
treated a more realistic model where the collective motion was coupled
self-consistently to the single-particle motion and found the coexistence of
undamped collective motion and chaotic single-particle dynamics. This
observation is interesting with view on low-lying collective excitations but
cannot apply to the case of the giant dipole resonance. Furthermore, the
classical models discussed in the literature suffer from the absence of a
two-body force and rotational symmetry. In this work we are particularly 
interested in the many-body aspects of the problem and the role of the 
two-body interaction opposed to the mean-field. To
this purpose we study a classical self-bound many-body system with the
following characteristics: two-body surface interaction; chaotic dynamics;
collective mode constructed from single-particle degrees of freedom.

Let us consider the model Hamiltonian
\be
\label{ham}
H=\sum_{i=1}^N{\vec{p}_i^2\over 2m} + \sum_{i<j}V(|\vec{r}_i-\vec{r}_j|),
\ee
where $\vec{r}_i$ is a two--dimensional position vector of the $i$-th nucleon
and $\vec{p}_i$ is its conjugate momentum. The interaction is given by
\ba
\label{int}
V(r)=\left\{
     \begin{array}{ll}
     0 & \mbox{for $r<a$}, \\
     \infty & \mbox{for $r\ge a$}.
     \end{array}
     \right.
\ea
The particles thus move freely within a convex billiard in $2N$-dimensional
configuration space and undergo specular reflection at the wall. This
corresponds to the basic picture of nucleon motion in a nucleus: they move in a
common flat-bottom potential, and interact mainly at the surface. The
Hamiltonian~(\ref{ham},\ref{int}) achieves these features while retaining the
two-body interaction.  Total momentum $\vec{P}$, energy $E$ and angular
momentum $L$ are conserved.  Classical phase space structure is independent of
energy since a scaling $\vec{p}_i\to\alpha\vec{p_i}$ simply leads to a
rescaling of $E\to\alpha^2 E, \vec{P}\to\alpha\vec{P}$, and $L\to\alpha L$.  We
set $m=a=\hbar=1$. Thus, energies, momenta and times are measured in units of
$\hbar^2/ma^2$, $\hbar/a$ and $ma^2/\hbar$, respectively
\footnote{We introduced $\hbar$ to set an energy scale.}.
In what follows we fix the total
momentum $\vec{P}=0$, the angular momentum $L=0$ and set the energy $E=N$.

The $N=2$ system is integrable and chaotic dynamics may exist for $N\ge 3$. To
study the classical dynamics of the $N$-body system we compute the Lyapunov
exponent using the tangent map \cite{Sieber}.  The time evolution of the
$N$-body system is done efficiently by organizing collision times in a
partially ordered tree and requires an effort ${\cal{O}}(N\ln{N})$
\cite{Prosen}. We draw several initial conditions for fixed $N$ at random from
the $(E=N,\vec{P}=0,L=0)$-shell and follow their time evolution for several
hundred Lyapunov times. This ensures good convergence of the Lyapunov
exponents. The results of several runs are listed in Table~\ref{tab1} (mean
values and RMS deviations). For $N=3$ we traced $7\times10^4$ trajectories, all
of them being unstable with respect to small initial deviations. Thus, we
expect more than 99\% of phase space to be chaotic. For larger $N$ we traced
fewer trajectories and have less statistics.  However, {\it all} followed
trajectories have positive Lyapunov exponents.  This suggests that the dynamics
of the $N$-body system is dominantly chaotic.  We note that the $N$-body system
possesses marginally stable orbits corresponding to configurations where $N-2$
particles are at rest.  However, these configurations are of zero measure in
phase space.  The reliability of the numerical evolution was checked by
comparing forward with backward evolution. Moreover, total energy, total
momentum and angular momentum were conserved to high accuracy.  As a further
check we computed the Lyapunov exponents using an alternative method
\cite{Bennetin} and found good agreement with the results displayed in
Table~\ref{tab1}.

We next consider the time evolution of collective motion. To this purpose we
define a set of initial conditions (i.e. phase space points) of the $N$-body
system that lead to collective motion corresponding to the giant dipole
resonance. In passing we mention that one could also use collective coordinates
introduced by Zickendraht \cite{Zickendraht} or consider motion close to
invariant manifolds of the rotationally symmetric many-body system \cite{PSW}.
Let us draw the momenta $\vec{p}_i$ at random from uniform probability
distributions that vanish outside of the domains $(p_x,p_y)\in[-\Delta
q_x,\Delta q_x]\times[\pm q-\Delta q_y,\pm q+\Delta q_y]$ , with $+$ or $-$
sign for particles $i=1,\ldots,N/2$ (e.g. protons) and particles
$i=N/2+1,\ldots,N$ (e.g. neutrons), respectively.  We rescale the energy to
$E=N$.  The initial positions $\vec{r}_i, i=1,\ldots,N-1$ are drawn at random
to lie inside a circle of radius $a/2$; the position of the $N$th particle is
chosen such that the total angular momentum vanishes; the center of mass is
subtracted.  For $\Delta q_x\approx \Delta q_y\ll q\approx\sqrt{2mE/N}$ one
obtains initial conditions that correspond to the motion of all protons against
all neutrons and therefore may be identified with the dipole giant
resonance\footnote{Our choice of zero angular momentum is justified in the
framework of classical mechanics.}.  In what follows we use $\Delta q_x=\Delta
q_y\approx q/10$.

\begin{figure}
  \begin{center}
    \leavevmode
    \parbox{0.9\textwidth}
           {\psfig{file=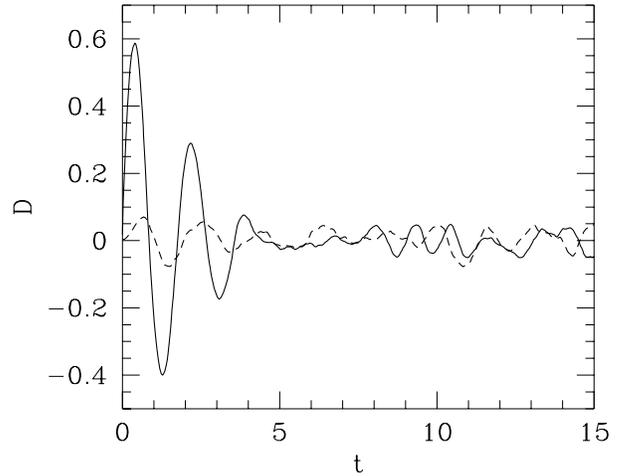,width=0.45\textwidth,angle=0}}
  \end{center}
\protect\caption{Time evolution of the 
dipole-moment~(\protect\ref{dipole}) for a system of $N=220$ nucleons
at energy $E=N$. (Full line: longitudinal $D_y$-component; dashed line:
transversal $D_x$-component.) Time and dipole-moment are given in units 
of $ma^2/\hbar$ and $a$, respectively.}  
\label{fig1}
\end{figure}

We are interested the time evolution of the dipole-moment
\be
\label{dipole}
\vec{D}={2\over N}\sum_{i=1}^{N/2} (\vec{r}_i-\vec{r}_{N-i}).  
\ee
Fig.~\ref{fig1} shows the time dependence of the dipole-moment for a system of
size $N=220$ and one collective initial condition.  The longitudinal component
$D_y$ exhibits a few damped oscillations that eventually turn into erratic
fluctuations of small amplitude. On the same time scale the transverse
component $D_x$ increases in amplitude from approximately zero and turns into
similar erratic fluctuations. These fluctuations decrease with increasing
system size $N$, indicating that they are of statistical nature.  The first two
periods of the time evolution can be fitted to an exponentially damped harmonic
oscillation yielding the period $\tau$ and the damping rate $\gamma$. The
results of several runs (mean values and RMS deviations are listed in
Table~\ref{tab2}.

\begin{figure}
  \begin{center}
    \leavevmode
    \parbox{0.9\textwidth}
           {\psfig{file=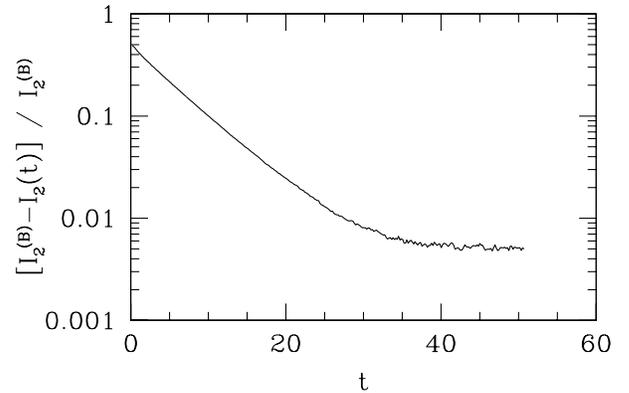,width=0.45\textwidth,angle=0}}
  \end{center}
\protect\caption{Second moment $I_2(t)$ of the single-particle energy 
compared to its equilibrium value $I_2^{(B)}$ for $N=220$. 
Time in units of $ma^2/\hbar$.}  
\label{fig2}
\end{figure}

\begin{figure}
  \begin{center}
    \leavevmode
    \parbox{0.9\textwidth}
           {\psfig{file=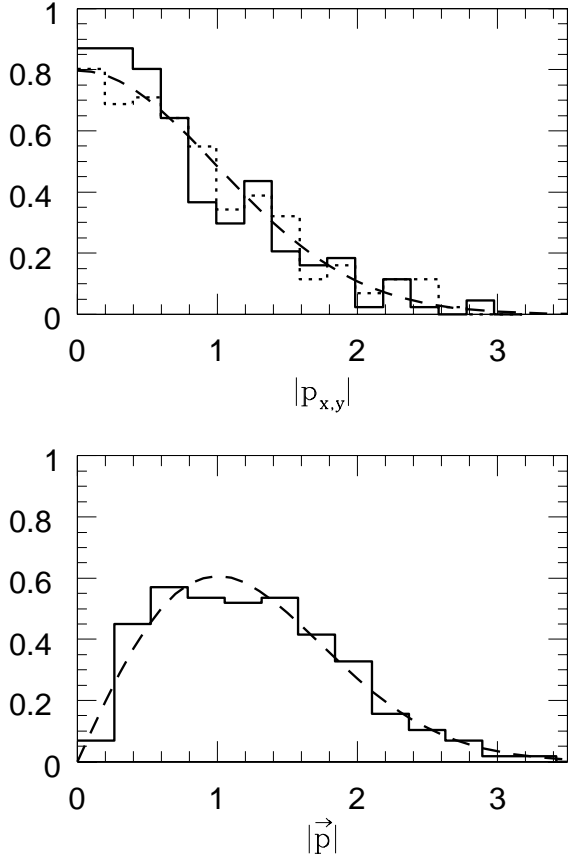,width=0.45\textwidth,angle=0}}
  \end{center}
\protect\caption{Top: Normalized momentum distributions of single 
particle momenta $|p_x|$ (top: full line), $|p_y|$ (top: dotted line) and 
$|\vec{p}|$ (bottom: full line) 
after the equilibration of the collective motion compared to Maxwell-Boltzmann 
distributions with temperature $kT=E/N=1$ (dashed lines). 
A system of $N=220$ nucleons is considered; momenta are given in units of 
$\hbar/a$.}
\label{fig3}
\end{figure}

For the equilibration we study the distribution of single-particle energies
$\epsilon_j, j=1\ldots N$ during the time evolution of the dipole mode. The
moments $I_\nu\equiv N^{-1}\sum_{j=1}^N\epsilon_j^\nu$ of this distribution may
be compared with the corresponding equilibrium values obtained from a Boltzmann
distribution $\rho^{(B)}(\epsilon)=1/[kT\exp{(\epsilon/kT)}]$ at a temperature
$kT=E/N=1$, i.e. $I_\nu^{(B)}=\nu !(kT)^\nu$.  Fig.~\ref{fig2} shows the first
nontrivial moment $I_2$ averaged over many runs to reduce statistical
fluctuations. This moment approaches its equilibrium value exponentially fast
at a rate $\alpha$ (listed in Table~\ref{tab2}). The saturation at long times
seems to be due to the finite number of particles. Once the system has
equilibrated, the single-particle momenta are Maxwell-Boltzmann distributed
(See Fig.~\ref{fig3}). We also computed the Lyapunov exponent $\lambda$ of the
many-body trajectory corresponding to the collective dipole motion and listed
the results (mean values and RMS deviations) of several runs in
Table~\ref{tab2}.  The Lyapunov exponents agree well with the previously
computed ones listed in Table~\ref{tab1}. This shows that the dynamics of the
many-body system is also chaotic in the region of phase space that corresponds
to the collective motion.  Table~\ref{tab2} displays a hierarchy of well
separated rates $\lambda > \gamma >\alpha$ with $\lambda\gg\alpha$. This
hierarchy is particularly pronounced for large $N$ since $\lambda$ ($\alpha$)
increases (decreases) with increasing $N$. The damping and equilibration rates
differ for the following reason. While the dipole mode is damped out once the
single-particle momentum distribution exhibits spherical symmetry,
equilibration requires also a considerable change in the radial momentum
distribution. Note that the dipole moment $\vec{D}$ and the moments $I_\nu$ of
the single-particle energy distribution effectively test the angular and radial
momentum distribution, respectively. Our results show that a generic self-bound
many-body system with two-body interactions displays a fast equilibration in
the angular components opposed to a slow equilibration in the radial components
of the single-particle momenta. This observation is also of interest with view 
on heavy ion collisions. 

\begin{figure}
  \begin{center}
    \leavevmode
    \parbox{0.9\textwidth}
           {\psfig{file=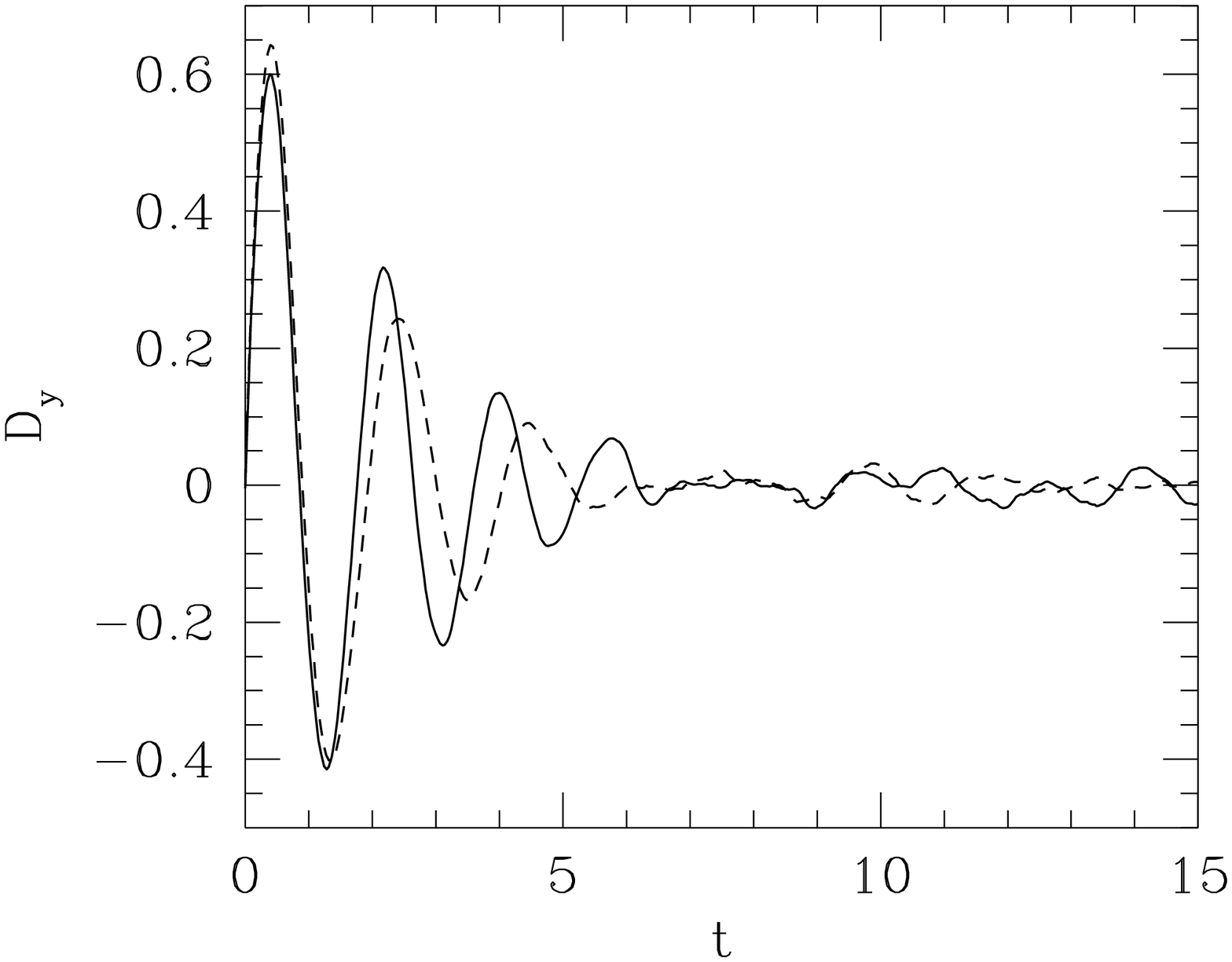,width=0.45\textwidth,angle=0}}
  \end{center}
\protect\caption{Time evolution of the 
longitudinal dipole-moments $D_y$ for a system of $N=500$ nucleons
at energy $E=N$. (Full line: system with two-body interaction; dashed line:
mean-field model.) Time and dipole-moment are given in units 
of $ma^2/\hbar$ and $a$, respectively.}  
\label{fig4}
\end{figure}

To examine the role of the two-body interaction more closely, we consider a
system of $N$ independent particles moving inside a circular billiard of
diameter $a$. This system is integrable since single-particle energies and
angular momenta are conserved quantities. It can be taken as a mean-field
approximation of our Hamiltonian \cite{BPR} and is motivated by the observation
that the surface of the interacting many body-system (i.e. the points in
two-dimensional configuration space where interactions occur) becomes sharper
with increasing numbers of particles and may roughly be approximated by a
circle of diameter $a$. As before, we take initial conditions at random from
the region that corresponds to collective motion and compute the dipole-moment
(\ref{dipole}) as a function of time.  Fig.~\ref{fig4} compares the evolution
of the longitudinal dipole-moment $D_y$ obtained for the mean-field model with
the result obtained for the interacting many-body system.  One observes damped
oscillations with a period and a damping rate that are slightly larger than for
the interacting system. The mean-field Hamiltonian therefore captures the
damping of collective motion quite accurately. In the integrable system,
damping is due to the dephasing of single-particle oscillations that is induced
by the collisions with the surface, i.e. the directions of single-particle
momenta get randomized on a time scale $1/\gamma$. The vadility of the
mean-field picture is also confirmed by the observation that the periods and
damping rates display no significant $N$-dependence. This is, however,
different for the Lyapunov exponents and equilibration rates. These do depend
on $N$ through the presence of the two-body interaction. Note that our
observation of $N$-independent frequencies $\omega=2\pi/\tau$ and damping rates
$\gamma$ leads to a simple scaling law. Keeping the average single-particle
energies $E/N$ fixed and scaling the diameter of the $N$-body system as
$a\propto N^{1/3}$ yields $\omega,\gamma\propto N^{-1/3}$. These results are in
semi-quantitative agreement with experimental data \cite{Speth},
i.e. $\hbar\omega\propto N^{-1/3}$ in heavy nuclei, and $\gamma\propto
N^{-[1/3\ldots 2/3]}$.

We also considered initial conditions of the interacting many-body system with
larger momentum spread $\Delta q_x\approx \Delta q_y\approx q$, i.e.  the
initial momentum distribution is closer to an equilibrium distribution.  This
situation is closer to the perturbative excitation of the nuclear giant
resonance.  As expected, one finds a collective oscillation of smaller
amplitude and comparable period, showing that the observed phenomena are
robust. The determination of the damping rate is, however, difficult since
amplitide and statistical fluctuations are roughly of the same magnitude.

In summary, we have considered damping and equilibration of collective motion
in a self-bound $N$-body systems with two-body (surface) interactions. This
system is predominantly chaotic and exhibits damped collective motion that
leads to equilibration. There is a hierarchy of three well separated time
scales starting with the onset of classical chaos at short times, damping of
the dipole mode at intermediate times and equilibration at long times,
respectively. The damping is mainly due to dephasing and may be understood
quite accurately in a mean-field picture of non-interacting
particles. Consequently, periods and damping rates show no significant
$N$-dependence. Equilibration, however, requires the randomization of the
magnitudes of single-particle momenta and a two-body interaction.
Lyapunov exponents and equilibration rates depend mildly on $N$. The presented
model exhibits a rather homogeneous phase space structure and the main
characteristics of self-bound many-body systems like atomic nuclei. Our results
show that it captures important features of collective motion in such systems.

I thank G. Bertsch, F. Leyvraz, T. Prosen, S. Reddy, P.-G. Reinhard and
T. Seligman for useful discussions. The hospitality of the Centro Internacional
de Ciencias, Cuernavaca, Mexico, is gratefully acknowledged.  This work was
supported by the Department of Energy under grant DOE/ER/40561.

\begin{table}
\begin{tabular}{|c||c|c|c|c|c|c|}
$ N$             & 3     & 100   & 150   & 220  & 330  & 500  \\\hline
$\lambda $       & 0.963 & 1.506 & 1.561 & 1.62 & 1.67 & 1.75 \\
$\Delta\lambda $ & 0.004 & 0.017 & 0.020 & 0.02 & 0.02 & 0.02 \\
Runs             & 70000 & 100   & 100   & 50   & 30   & 10   \\
\end{tabular}
\protect\caption{Lyapunov exponents (mean values $\lambda$ and RMS 
deviations $\Delta\lambda$) for
$N$-body systems with unit single-particle energy $E/N=1$. Results are
obtained from the listed number of runs, $\lambda$ is given in units of
$\hbar/ma^2$.} 
\label{tab1}
\end{table}

\begin{table}
\begin{tabular}{|c||c|c|c|c|c|}
$ N$     & 100            & 150            & 220            & 330            & 500  \\\hline
$\lambda$& 1.52 $\pm$0.02 & 1.57 $\pm$0.01 & 1.61 $\pm$0.03 & 1.67 $\pm$0.02 & 1.73 $\pm$0.02 \\
$\gamma$ & 0.39 $\pm$0.05 & 0.38 $\pm$0.08 & 0.42 $\pm$0.06 & 0.39 $\pm$0.03 & 0.37 $\pm$0.04 \\
$\alpha$ & 0.21           & 0.19           & 0.15           &    0.13        & 0.10           \\
$\tau$   & 1.77 $\pm$0.05 & 1.77 $\pm$0.04 & 1.76 $\pm$0.03 & 1.77 $\pm$0.02 & 1.77 $\pm$0.01 \\
\end{tabular}
\protect\caption{Damping rates $\gamma$, periods $\tau$, equilibration rates 
$\alpha$ and Lyapunov exponents $\lambda$ of the collective mode. 
($\gamma$,$\tau$,$\lambda$: Mean values and RMS deviations are 
obtained from ten runs for each listed $N$-body system. $\alpha$: exponent
obtained from average of many runs.) The single-particle
energy is $E/N=1$; $\tau,1/\lambda,1/\gamma$ and $1/\alpha$ are given in units of 
$ma^2/\hbar$.} 
\label{tab2}
\end{table}

\begin{references}  
\bibitem{BoWei} 
O. Bohigas and H. A. Weidenm\"uller,
Ann. Rev. Nucl. Part. Sci. {\bf 38} (1984) 421

\bibitem{Zele} 
V. Zelevinsky,
Ann. Rev. Nucl. Part. Sci. {\bf 46} (1996) 237

\bibitem{GMW} 
T. Guhr, A. M\"uller-Groeling, and H. A. Weidenm\"uller,
Phys. Rep. {\bf 299} (1998) 189 

\bibitem{Koonin}
S. E. Koonin and J. Randrup,
Nucl. Phys. A {\bf 289} (1977) 475

\bibitem{Blocki78}
J. Blocki, Y. Boneh, J. R. Nix, J. Randrup, M.Robel, A. J. Sierk, and W. J. Swiatecki,
Ann. Phys. {\bf 113} (1978) 330

\bibitem{Speth}
J. Speth and A. van der Woude,
Rep. Prog. Phys. {\bf 44} (1981) 719

\bibitem{Bertsch}
G. F. Bertsch, R. F. Bortignon, and R. A. Broglia,
Rev. Mod. Phys. {\bf 55} (1983) 287

\bibitem{Wilkinson}
M. Wilkinson,
J. Phys. A {\bf 21} (1988) 4021

\bibitem{Abul}
A.Y. Abul-Magd and H. A. Weidenm\"uller,
\pl B {\bf 261} (1991) 207 

\bibitem{Blocki93}
J. Blocki, J. J. Shi, and W. J. Swiatecki,
Nucl. Phys. {\bf A554} (1993) 387

\bibitem{Aurel}
A. Bulgac, G. Do Dang, and D. Kusnezov, 
Phys. Rep. {\bf 264} (1996) 67

\bibitem{Hofmann} 
H. Hofmann, 
Phys. Rep. {\bf 284} (1997) 137 

\bibitem{Baldo}
M. Baldo, G. F. Burgio, A. Rapisarda, and P. Schuck,
\prc {\bf 58} (1998) 2821

\bibitem{Bauer}
W. Bauer, D. McGrew, V. Zelevinsky, and P. Schuck,
\prl {\bf 72} (1994) 3771

\bibitem{Sieber} 
M. Sieber and F. Steiner,
Physica {\bf D44} (1990) 248

\bibitem{Prosen}
T. Prosen, private communication.

\bibitem{Bennetin} 
G. Bennetin, L. Galgani, and J. M. Strelcyn,
\pra {\bf 14} (1976) 2338 

\bibitem{Zickendraht}
W. Zickendraht,
J. Math. Phys. {\bf 12} (1971) 1663

\bibitem{PSW}
T. Papenbrock, T. H. Seligman, and H. A. Weidenm\"uller,
\prl {\bf 80} (1998) 3057

\bibitem{BPR}
G. F. Bertsch, T. Papenbrock, and S. Reddy,
e-print nucl-th/9906054 

\end{references}
\end{document}